\documentclass{ws-mpla}
\newcommand{\lk}{\left( }
\newcommand{\rk}{\right)}
\newcommand{\ltk}{\left\{ }
\newcommand{\rtk}{ \right\} }
\newcommand{\ldk}{\left[ }
\newcommand{\rdk}{ \right] }

\begin{document}

\markboth{KANABU NAWA, HIDEO SUGANUMA and TORU KOJO}
{BARYONS WITH HOLOGRAPHY}

\catchline{}{}{}{}{}

\title{BARYONS WITH HOLOGRAPHY
}

\author{\footnotesize KANABU NAWA}
\address{Research Center for Nuclear Physics (RCNP), Osaka University, \\
Mihogaoka 10-1, Ibaraki, Osaka 567-0047, Japan\\
nawa@rcnp.osaka-u.ac.jp}

\author{HIDEO SUGANUMA and TORU KOJO}

\address{Department of Physics, Kyoto University, Sakyo, Kyoto 606-8502, Japan}

\maketitle


\begin{abstract}
We perform the first study of baryons in holographic QCD with
${\rm D}4/{\rm D}8/\overline{{\rm D}8}$ multi-${\rm D}$-brane system
in type IIA superstring theory.
The baryon is described as a chiral soliton solution 
in the four-dimensional meson effective action derived from 
holographic QCD. 
We also present a brief review of the holographic model
from the viewpoints of recent hadron physics and QCD phenomenologies.
\end{abstract}

\ccode{PACS Nos.: 11.25.Uv, 12.38.-t, 12.39.Dc, 12.39.Fe}

\section{Review of Holography}	

Holography is a new concept of the superstring theory
based on the duality between gauge and supergravity theories,
proposed by Maldacena in 1997.\cite{Mal}
The superstring theory with conformal and Lorentz
invariance has ten-dimensional space-time to avoid anomalies.
Its elemental degrees of freedom are one-dimensional open or closed
strings.
The fluctuation modes of these strings are expected to give the
elementary particle degrees of freedom as in the standard model.
One can also find a soliton of strings as the ($p$+1)-dimensional 
membrane composed by the condensed strings, called ${\rm D}_p$-brane.\cite{Poli}
In particular, ${\rm D}_p$-brane has two important aspects as follows.

First, a ($p$+1)-dimensional gauge theory appears {\it on} ${\rm D}_p$-brane. 
Here, an open string with the two edges on the ${\rm D}_p$-brane has 
ten independent fluctuation modes; scalar fields $\Phi_{i=(p+1)\sim 9}$
and the other vector-like fields ${\cal A}_{\mu=0\sim p}$.
${\cal A}_{\mu=0\sim p}$ have an index $\mu$ on ${\rm D}_p$-brane extending to ($p$+1)-dimensional space-time,
and ${\cal A}_{\mu}$ fields can be regarded as gauge fields on the ${\rm D}_p$-brane.
In this sense, ($p$+1)-dimensional gauge theory appears {\it on} ${\rm D}_p$-brane. 

Second, $\{$($p$+{\it 1})+1$\}$-dimensional supergravity appears {\it around} 
the ${\rm D}_p$-brane.
In fact, ${\rm D}_p$-brane is found to be a strong 
gravitational system in ten-dimensional space-time 
as a black hole or a black brane,\cite{Poli} giving nontrivial
curvatures around it.
Due to the geometrical symmetry, the angle coordinates 
around ${\rm D}_p$-brane become trivial and can be integrated out in the 
action of ${\rm D}_p$-brane.
Then, there eventually remains just one radial dimension
with nontrivial curvature, indicating the existence of gravity.
In this sense, $\{$($p$+{\it 1})+1$\}$-dimensional supergravity appears {\it around}
${\rm D}_p$-brane. (The italic ``{\it 1}'' denotes the radial dimension around
${\rm D}_p$-brane.)

Now we discuss the holography.
Holography indicates the equivalence between the ($p$+1)-dimensional gauge theory
without gravity
{\it on} ${\rm D}_p$-brane
and the $\{$($p$+{\it 1})+1$\}$-dimensional supergravity {\it around} ${\rm D}_p$-brane. 
Dimensions are different with each other to give the concept of holography.
In particular, the ($p$+1)-dimensional gauge theory projected on the surface of 
${\rm D}_p$-brane
is often called as the ``hologram''
of the $\{$($p$+{\it 1})+1$\}$-dimensional supergravity.

The most essential property of holography is the existence of
strong-weak duality (S-duality) between the gauge theory and the supergravity:
couplings are transversely related with each other.
Therefore, if one wants to analyze the non-perturbative
(strong-coupling) aspects of one side,
one can attack from the other dual side only with 
the tree-level (weak-coupling) calculations.
Consequently, this new concept of holography is expected to be a powerful tool 
for both sides of the gauge theory in hadron physics and 
the gravitational theory in astrophysics.\cite{gra1,gra2}

\section{Holographic QCD}

If one succeeds in constructing QCD with quarks and 
gluons on a special configuration of ${\rm D}$-branes, 
one can attack the non-perturbative aspects of QCD
from the tree-level dual supergravity side.
This is the strategy of holographic QCD.

In 2005, Sakai and Sugimoto succeeded in constructing massless QCD
on 
${\rm D}4/{\rm D}8/\overline{{\rm D}8}$ multi-${\rm D}$-brane system.\cite{SS}
$\overline{{\rm D}8}$-brane has opposite chirality relative to 
${\rm D}8$-brane to introduce the concept of chiral symmetry in this
model.
Then, by using the concept of holography,
non-perturbative aspects of massless QCD can be analyzed 
by the tree-level calculation in the dual  $\{$(4+{\it 1})+1$\}$-dimensional 
supergravity side, called holographic QCD.
In fact, this theory describes mesons as color-singlet low-energy objects, 
and reproduces meson mass spectra in coincidence with
experimental data, and also other many traditional meson phenomenologies like
vector-meson dominance, KSRF relation, GSW model and so on.\cite{SS}
In this sense, holographic QCD can be regarded as the ``unified meson
theory'' based on QCD.

In this construction, however, $\{$(4+{\it 1})+1$\}$-dimensional 
classical supergravity is dual with large-$N_c$ QCD. 
Then, as a general property of large-$N_c$ QCD,
another important color-singlet objects as {\it baryons}
do not directly appear in the large-$N_c$ limit.\cite{tH}
In fact, it is non-trivial to describe baryons
in the large-$N_c$ holographic QCD.

\section{Baryons as Brane-induced Skyrmions in Holographic QCD}

In order to describe baryons in holographic QCD,
we introduce the concept of chiral solitons (Skyrmions) 
in the large-$N_c$ holographic model.
The Skyrme model is first proposed in 1961,\cite{Sky}
describing a baryon as a topological chiral soliton of the pion field, 
which is the Nambu-Goldstone boson relating to spontaneous chiral-symmetry breaking.
Here, the stability of the chiral soliton is known to be sensitive 
to the four-derivative terms of pion fields.
Chiral and Lorentz symmetries allows three candidates as four-derivative terms:
\begin{equation}
{\rm tr}[L_\mu, L_\nu]^2,\hspace{4mm}{\rm tr}\{L_\mu, L_\nu\}^2,\hspace{4mm}
{\rm tr}(\partial_\mu  L_\nu)^2, 
\label{deri1}
\end{equation}
where $L_\mu$ is one-form of pion fields, $L_\mu=(1/i)U^\dagger\partial_\mu U$.
The first term in Eq.(\ref{deri1}) is the Skyrme term.
The other two candidates in Eq.(\ref{deri1}) are known to
give instability of the Skyrmion,\cite{ZB} but these two cannot be excluded by the symmetry arguments. 

As a remarkable fact, in the leading order of holographic QCD, 
only the Skyrme term in Eq.(\ref{deri1}) appears. 
This can be understood from the fact that 
the leading order of
the effective action (DBI action) of ${\rm D}8$-brane
includes only ``two time-derivatives'' at most, which gives a severe restriction
on the possible terms in the effective meson theory.
Thus, the topological picture for baryons is derived from QCD and 
seems to be justified ingeniously in this framework with the superstring theory. 

\begin{figure}[t]
\centerline{\psfig{file=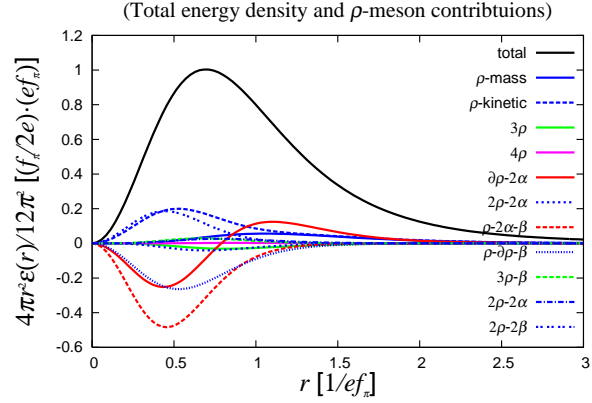,width=2.9in}}
\vspace*{8pt}
\caption{Total energy density of a baryon 
and each contribution of $\rho$-meson interaction terms in the meson effective action derived
from holographic QCD. (For details, see Ref.9. Although all the calculations are correct, 
there is a typo of an overall factor on vertical values in Figs.6-8 of Ref.9.)
\label{fig1}
}
\end{figure}

We investigate the hedgehog configuration for pion fields $U({\bf x})$ 
and $\rho$ meson fields $\rho_\mu ({\bf x})$ 
in the meson effective action derived from holographic QCD as 
\begin{eqnarray}
U^{\star}({\bf x})=e^{i\tau_a \hat{x}_a F(r)}, 
\hspace{5mm}
\rho^{\star}_{0}({\bf x})=0, \hspace{5mm} \rho^{\star}_{i}({\bf x})=\rho^{\star}_{i a}({\bf x}){\tau_a}/2
                                               =\ltk \varepsilon_{i a
                                               b}\hat{x}_b\tilde{G}(r)\rtk
                                               \tau_a.
\label{WYTP}
\end{eqnarray}
$F(r)$ ($r \equiv |{\bf x}|$) is a dimensionless function 
with boundary conditions $F(0)=\pi$ and $F(\infty)=0$, giving
topological charge equal to unity as a unit baryon number.
With the hedgehog configuration Ansatz (\ref{WYTP}), we uniqeuly derive 
the energy density $\varepsilon [F(r), \tilde{G}(r)]$ for the brane-induced Skyrmion 
including interaction terms of pions and $\rho$ mesons 
from holographic QCD as follows:
\begin{eqnarray}
& &r^2\varepsilon [F(r), \tilde{G}(r)] =
             \frac{f_\pi^2}{4}\ldk 2\lk r^2 F^{'2}+2\sin^2F \rk\rdk
          + \frac{1}{32e^2}\ldk 16 \sin^2F \lk 2 F^{'2}+\frac{\sin^2F}{r^2}\rk\rdk 
\nonumber\\
         &+& \frac{1}{2}%
             \ldk 8\ltk
             3\tilde{G}^2+2 r \tilde{G}(\tilde{G}^{'}) +r^2\tilde{G}^{'2}\rtk\rdk
          + m_{\rho}^2%
             \ldk 4 r^2 \tilde{G}^2\rdk
          - g_{3\rho}  %
             \ldk 16 r \tilde{G}^3 \rdk
          + \frac{1}{2}g_{4\rho}%
             \ldk 16 r^2 \tilde{G}^4 \rdk
\nonumber\\
         &+&  g_1 %
             \ldk 16 \ltk F^{'}\sin F \cdot\lk
             \tilde{G}+ r \tilde{G}^{'}\rk%
             +\sin^2F\cdot\tilde{G}/r\rtk\rdk
          - g_2%
             \ldk 16 \sin^2F\cdot\tilde{G}^2 \rdk
\nonumber\\
         &-& g_3 %
             \ldk 16 \sin^2F\cdot\lk 1-\cos F\rk \tilde{G}/r \rdk 
          - g_4%
             \ldk 16 \lk 1-\cos F\rk\tilde{G}^2\rdk   
          + g_5 %
             \ldk 16 r \lk 1-\cos F \rk \tilde{G}^3\rdk
\nonumber\\
          &+& g_6%
             \ldk 16 r^2 F^{'2}\tilde{G}^2\rdk
          + g_7%
             \ldk 8 \lk 1-\cos F\rk^2\tilde{G}^2\rdk.
\label{energy_dense_An}
\end{eqnarray}  
%
All the coupling constants 
($e$, $g_{3\rho}$, $g_{4\rho}$, $g_{1\sim 7}$) in Eq.(\ref{energy_dense_An}) are uniquely
determined by just two experimental inputs as
$f_\pi=92.4\mbox{MeV}$ and $m_\rho=776\mbox{MeV}$, which is a remarkable
consequence of holographic framework.
We numerically obtain the stable chiral soliton solution 
as a baryon in holographic QCD, which basically supports the ``Skyrme soliton picture'' for baryons 
with corrections by $\rho$ mesons.
Figure~\ref{fig1} shows the total energy density of a baryon and
each contribution of $\rho$ meson interaction terms derived from
holographic QCD. This result indicates the active contribution of $\rho$ mesons in 
the deeper interior region of baryons, and such a ``$\rho$ mesonic parton'' in the soliton 
may lead to a new picture for baryons from holographic framework. 

\section{Summary and Concluding Remarks}
We have performed the first study of baryons as chiral solitons in holographic QCD with
${\rm D}4/{\rm D}8/\overline{{\rm D}8}$ multi-${\rm D}$-brane system.\cite{NSK1,NSK2} 
We have found that the holographic QCD has a stable hedgehog soliton solution as a baryon, 
which is basically described as the Skyrme soliton of pions including $\rho$ mesons in the interior region.

Just after our study, similar studies of baryons were done using instantons in the
five-dimensional gauge theory in holographic framework.\cite{ins1,ins2}
It is also interesting to compare chiral solitonic baryons in the holographic QCD 
with other-type construction of stringy hadrons in a holographic framework.\cite{TB05}

From more general point of view, ``holography'' is a powerful concept which links 
QCD-hadron physics and blackhole-astrophysics by virtue of its S-duality. 
Large applications of this new concept are much expected in near future.

\end{document}